\documentclass[fleqn,usenatbib]{mnras}
\usepackage{multirow}
\usepackage{newtxtext,newtxmath}
\usepackage{titlesec}

\usepackage[T1]{fontenc}
\DeclareRobustCommand{\VAN}[3]{#2}
\let\VANthebibliography\thebibliography
\def\thebibliography{\DeclareRobustCommand{\VAN}[3]{##3}\VANthebibliography}

\usepackage[pagewise]{lineno}

\usepackage{graphicx}	
\usepackage{amsmath}	
\usepackage{comment}

\title[{\it NuSTAR} \& {\it NICER} view of RX~J0032.9-7348]{Broadband study of the SMC pulsar RX~J0032.9-7348 during its X-ray brightening in 2024 }

\author[Chhotaray et al.]{
Birendra Chhotaray,$^{1}$\thanks{E-mail: birendra@prl.res.in, rsbirendra786@gmail.com}
Gaurava K. Jaisawal,$^{2}$
Sachindra Naik,$^{1}$
Arghajit Jana$^{3}$
\\
$^{1}$Astronomy and Astrophysics Division, Physical Research Laboratory, Navrangpura, Ahmedabad - 380009, Gujarat, India\\
$^{2}$DTU Space, Technical University of Denmark, Elektrovej 327-328, DK-2800 Lyngby, Denmark\\
$^{3}$Instituto de Estudios Astrof\'{\i}sicos, Facultad de Ingenier\'{\i}a y Ciencias, Universidad Diego Portales, Av. Ej\'{e}rcito Libertador 441, Santiago, Chile
\\
}

\date{Accepted XXX. Received YYY; in original form ZZZ}

\pubyear{2024}

\begin{document}
\label{firstpage}
\pagerange{\pageref{firstpage}--\pageref{lastpage}}
\maketitle

\begin{abstract}
We present the results of the broadband timing and spectral analysis of the poorly understood SMC pulsar RX~J0032.9-7348 (= SXP 7.02) using {\it NuSTAR} and {\it NICER} observations during its X-ray brightening in 2024. Our timing analysis revealed a pulsation period of approximately 7.02~s in the X-ray light curve. The pulse profile obtained in the broad energy range is double-peaked and asymmetric in nature and shows moderate variation with the energy. An absorbed power-law model describes the 0.5-8 keV {\it NICER} spectra well. The 3-50 keV {\it NuSTAR} spectrum is best described with an absorbed power-law modified with a high-energy cutoff model. We find no evidence of iron or cyclotron line features in the energy spectrum. During our observation period, the 0.5-50 keV luminosity varies in the range of $\sim 8\times10^{36} - 4\times10^{37}$ erg~s$^{-1}$. We also discuss the dependence of spectral parameters on the rotational phase of the pulsar through phase-resolved spectroscopy. 
\end{abstract}


\section{Introduction}
Accretion-powered X-ray pulsars (XRPs) are magnetized neutron stars, which are part of X-ray binary (XRB) systems discovered in the 1970s \citep{1972ApJ...172L..79S}. The XRPs emit X-rays by accreting mass from their binary companion. These pulsars are characterized by intense magnetic fields, typically ranging from $\sim 10^{12}-10^{14}$ G, which direct the infalling material toward its magnetic poles, where most of the X-ray photons are generated. When the magnetic and spin axes are misaligned, the resulting emission is observed as pulsations from the neutron star~\citep{1973ApJ...179..585D,1983ApJ...270..711W}. The neutron star accretes matter through three different processes, depending on the physical properties and evolutionary stage of the companion star. These modes of accretion are: (i) Roche lobe overflow, (ii) stellar wind accretion, and (iii) accretion through the Be circumstellar disc.  More specifically, the latter two processes are prominent modes of mass transfers in high-mass X-ray binary systems, where a neutron star accretes from its massive optical companion~\citep{1976ApJ...204..555S,1995IAUS..163..271K}. 

The XRPs exhibit X-ray brightening phenomena or outbursts due to a sudden enhancement in the mass accretion rate to the neutron star. The material accreted onto the neutron star is decelerated either through gas-mediated or radiative shocks, depending on the accretion rate, before reaching the surface of the neutron star~\citep{1983ApJ...270..711W}. Near the polar regions, the interaction between matter and photons becomes the main source of X-ray radiation. Soft X-rays or seed photons, produced in the hot spots near the magnetic poles, are Compton up-scattered by the infalling plasma, leading to broadband X-ray emission~\citep{2007ApJ...654..435B}. The continuum spectra of these systems are typically described by a combination of a blackbody and a cutoff power-law continuum ~\citep{2006ApJ...647.1293N,2007ApJ...654..435B,2008ApJ...672..516N,2013A&A...551A...1R,2015MNRAS.448..620J}. Furthermore, accreting pulsars often show iron emission lines around 6.4~keV and cyclotron resonant scattering features (CRSF) in the 10–100 keV range, the latter offering a direct method to estimate the magnetic field strength of the pulsar~\citep{2015A&ARv..23....2W, 2016MNRAS.461L..97J, 2019A&A...622A..61S,2023MNRAS.518.5089C,2024MNRAS.534.2830C}.

Beyond the spectral properties, the accreting pulsars exhibit distinctive temporal behaviors. These include coherent pulsations associated with the spin period and spin-up episodes of the neutron star. The spin-up rate correlates with increased luminosity and is significantly observed in the case of the Be/X-ray binary (BeXRBs), where the neutron star accretes mass from the circumstellar disc of the Be star ~\citep{1973ApJ...184..271L,2011Ap&SS.332....1R}. The pulse profiles of these systems exhibit a change in shape with the mass accretion rate and energy~\citep{1989PASJ...41....1N, 2018ApJ...863....9W, 2023MNRAS.521.3951J}. Power density spectra (PDS) reveal sharp peaks corresponding to pulsations from the hot spot on the surface, and broader features linked to the quasi-periodic oscillations (QPOs) which are possibly associated with the motion of in-homogeneously distributed matter in the inner accretion disc \citep{1987ApJ...316..411V,2018ApJ...863....9W}. 

RX~J0032.9-7348 is an X-ray transient source discovered in the 0.1–2.4 keV energy range during the ROSAT survey of the Small Magellanic Cloud (SMC) conducted with ROSAT/PSPC between 1991 and 1993~\citep{1996A&A...312..919K}.  The source also exhibited minor variations in the X-ray intensity during that period. However, the nature of the source is still poorly understood. Recently, an X-ray brightening of RX~J0032.9-7348 was reported by the Wide-field X-ray Telescope (WXT) onboard the Einstein Probe (EP) mission on 27 October 2024 \citep{2024ATel16880....1J}. Furthermore, the EP observed the source with the Follow-up X-ray telescope (FXT) on 28 October 2024 two times at a gap of $\approx$6 hours. The source was detected at a flux level of $\approx$7.95$\times$10$^{-11}$ erg s$^{-1}$ cm$^{-2}$ in the $0.5-10$ keV energy range during the first observation, and at a similar flux level during second observation. XMM-Newton also observed the source on 02 November 2024, where timing analysis of the EPIC data discovered pulsed emission at 7.02~s \citep{2024ATel16901....1H}. Pulsation was discovered at a flux level of $\approx$4.2$\times$10$^{-11}$ erg s$^{-1}$ cm$^{-2}$ in the $0.2-10$ keV energy band~\citep{2024ATel16901....1H}. Additionally, Swift SMC Survey (S-CUBED) observed RX~J0032.9-7348 on 22 October 2024 and 11 November 2024 at a flux level of $\approx$1.46$\times$10$^{-11}$ erg s$^{-1}$ cm$^{-2}$ and $\approx$2.77$\times$10$^{-11}$ erg s$^{-1}$ cm$^{-2}$, respectively, in the $0.3-10$ keV energy range~\citep{2024ATel16900....1G}. Recently, \citet{2024ATel16904....1M} used the Southern African Large Telescope (SALT) for optical spectroscopic observations on 7 November 2024, aiming to identify the optical counterpart among two proposed candidates. However, the optical counterpart of RX~J0032.9-7348 remains unidentified. Their analysis of 23 years of OGLE I-band photometric data revealed no periodicity between 2 and 200 days, despite the expectation of a 20–30d orbital periodicity based on the $\approx$7 s pulsation from the neutron star, as per the Corbet diagram.

Following the reported X-ray brightening of the source, we proposed a Target of Opportunity (ToO) observation with the Nuclear Spectroscopic Telescope Array ({\it NuSTAR}), which was carried out on 17 November 2024 for an exposure of $\approx$25 ks. We also monitored the source with the Neutron star Interior Composition ExploreR ({\it NICER}) to investigate the timing and spectral properties of the source (Table~\ref{tab:nustar_log}). The preliminary results obtained from the {\it NuSTAR} observation are present in \citet{2024ATel16930....1C}. Here, for the first time, we present the detailed results obtained from the broadband timing and spectral analysis of RX~J0032.9-7348 using {\it NuSTAR} and {\it NICER} observations. The paper is structured as follows. Section~\ref{sec:2 X-ray} provides an overview of the observations and data reduction procedures applied to {\it NuSTAR} and {\it NICER} data. In Sections~\ref{sec:xray_timing} and \ref{sec:xray_spectral}, details of the results of X-ray timing and spectral analyses, respectively, are presented. Following this discussion, conclusions are presented in Sections~\ref{discussion} and ~\ref{conclusion}, respectively.

\begin{table}
	\centering
	\caption{Observation log of RX~J0032.9$-$7348}
	\label{tab:nustar_log}
	\begin{tabular}{lllll} 
		\hline
Observatory & Obs. ID & Date of Obs. (MJD)  &   Exposure (s)    \\ [2 pt]
\hline 
{\it NuSTAR} & 91001353002 & 2024-11-17 (60631)   & 24975 \\
{\it NICER} & 7205010102 & 2024-11-03 (60617)     & 1096 \\
{\it NICER} & 7205010113 & 2024-11-19 (60633)     & 481 \\
\hline
\end{tabular}
\end{table}

\begin{figure}
    \centering
    \includegraphics[trim={0 1cm 0 0.5cm},scale=0.3]{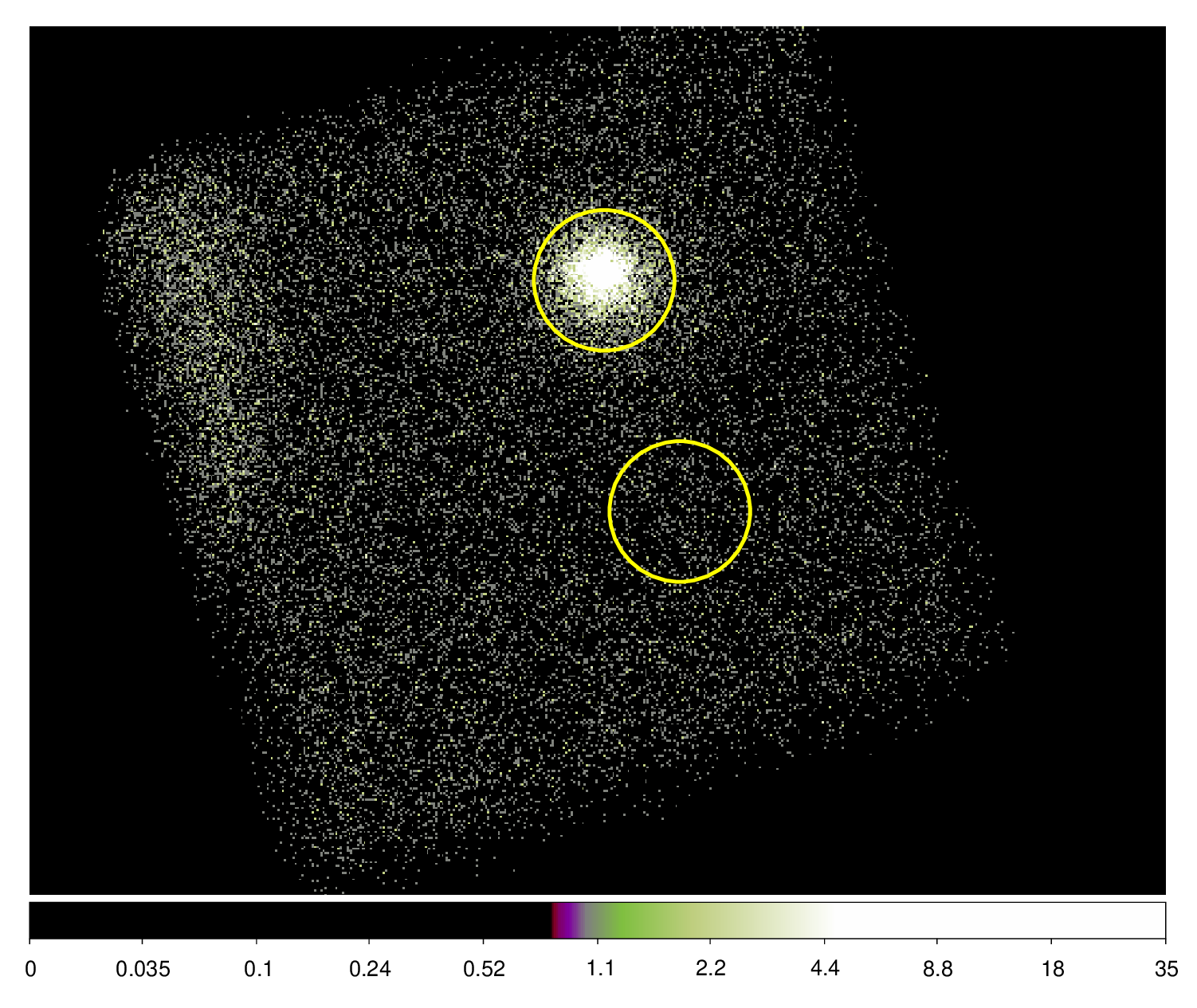}
    \caption{The source and background region selection using cleaned event file of the {\it NuSTAR} observation on MJD 60631.}
    \label{fig:ds9_rx}
\end{figure}

\begin{figure}
    \centering
    \includegraphics[trim={0 1cm 0 0.5cm},scale=0.35, angle=-90]{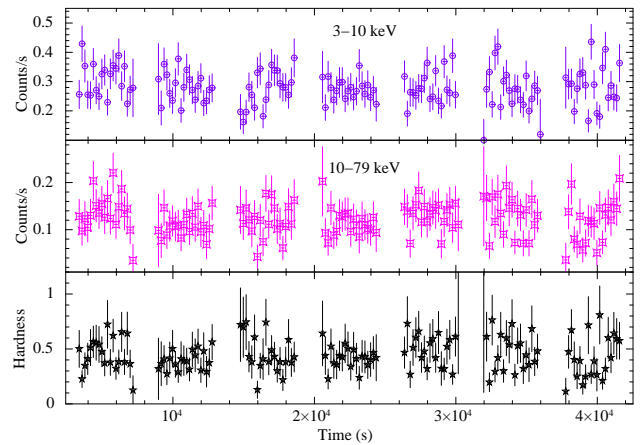}
    \caption{The top and middle panels display the {\it NuSTAR} light curves of RX~J0032.9-7348 in the 3–10 keV and 10–79 keV energy ranges, respectively. The bottom panel shows the hardness ratio, defined as the ratio of count rates in the 10–79 keV to 3–10 keV energy bands. The light curves are binned at 200 seconds. }
    \label{fig:lcurve_rx}
\end{figure}

\begin{figure*}
    \centering
    \includegraphics[trim={0 3.5cm 0 0.0cm},scale=1.4]{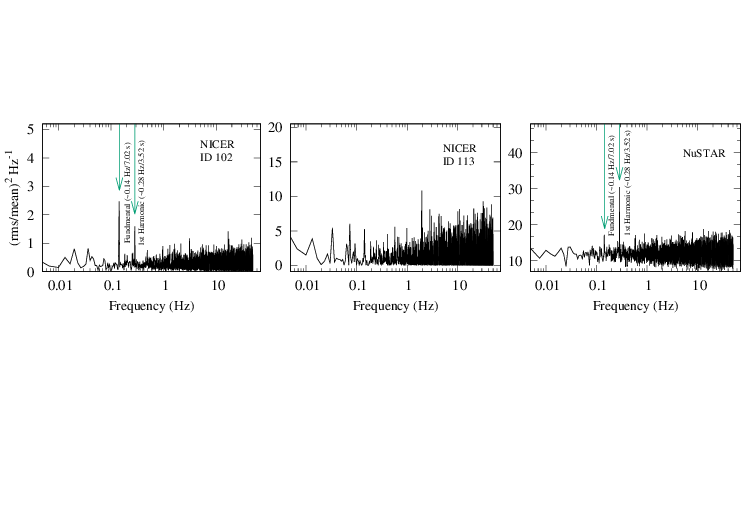}
    \caption{Power density spectrum (PDS) of RX~J0032.9-7348 obtained from {\it NICER} and {\it NuSTAR} observations. The frequency and period corresponding to the fundamental and 1st harmonic peak are annotated in the figure.  }
    \label{fig:powspec}
\end{figure*}

\section{Observation and Data Reduction}
\label{sec:2 X-ray}
\subsection{{\it NuSTAR}}
{\it NuSTAR} is a grazing incidence hard X-ray focusing telescope, sensitive in 3–79 keV energy range \citep{2013ApJ...770..103H}. It consists of two identical focal plane modules: Focal Plane Module A (FPMA) and Focal Plane Module B (FPMB), where data from each module are processed separately. RX~J0032.9-7348 was observed with {\it NuSTAR} on MJD~60631 (ID~91001353002) through a Target of Opportunity (ToO) observation (see Table~\ref{tab:nustar_log}). Data reduction was carried out using the HEASoft v.6.34 and the calibration files of version 20241104.  We followed the standard data processing routines where the {\tt NUPIPELINE} and {\tt NUPRODUCTS} tasks are executed to reprocess the unfiltered files to extract light curves and spectra. The source is detected at its known coordinates, RA (J2000) = 00:32:52.24 (8.21765d) and DEC (J2000) = -73:48:27.54 (-73.80765d), during the {\it NuSTAR} observation. For source and background products, the source and background regions were chosen using \texttt{ds9} software (see Figure~\ref{fig:ds9_rx}). The source region is centered on the source position, whereas the background region is chosen far away from the source position. Both the source and background regions are circular in shape with a radius of 80 arcsec.

\subsection{{\it NICER}}
{\it NICER}, launched in June 2017 and mounted on the International Space Station, is equipped with the X-ray Timing Instrument (XTI; \citealt{2016SPIE.9905E..1HG}), which operates in the 0.2–12 keV energy range. The XTI consists of 56 X-ray concentrator optics paired with silicon drift detectors, enabling non-imaging observations \citep{2016SPIE.9905E..1IP}. It provides high timing precision with a resolution of approximately 100 ns (rms) and a spectral resolution of around 85 eV at 1 keV. The field of view spans roughly 30 arcmin$^{2}$, and its effective area is about 1900 cm$^{2}$ at 1.5 keV, utilizing 52 active detectors. This study uses {\it NICER} observations of RX~J0032.9-7348 on MJD 60617 and 60633 (see Table~\ref{tab:nustar_log}). We use the {\tt nicerl2} pipeline available in \textsc{HEASoft} version 6.34 to process the unfiltered event data of {\it NICER}. The analysis is performed in the presence of gain and calibration database files of version \texttt{xti20240206}. We also applied barycentric correction using the solar system ephemeris \texttt{JPL-DE200} to correct for Earth and satellite motion during observations. Subsequently, we extracted light curves from each observation using the \textsc{XSELECT} task. We used \texttt{nicerl3-spect} task to generate spectrum, background estimate, and responses (ARF and RMF). The background spectrum is generated using the \texttt{3C50} model~\citep{2022AJ....163..130R}.

\section{Timing Analysis and Results}
\label{sec:xray_timing}
The X-ray timing analysis is performed using the {\it NuSTAR} light curve in the 3–79 keV energy range and the {\it NICER} light curves in the 0.5–10 keV energy range. The background-subtracted and barycenter-corrected light curves are generated at a bin size of 0.01 second. Correction due to binary motion is not applied as the orbital parameters for this system are not known. Figure~\ref{fig:lcurve_rx} displays the light curves of RX~J0032.9-7348 in the 3–10 keV and 10–79 keV energy ranges, along with the hardness ratio, defined as the ratio between the count rates in 10–79 keV and 3–10 keV ranges. No significant variation in the count rate or hardness ratio is seen during the {\it NuSTAR} observation. We also observed the same scenario in the case of {\it NICER} light curves. 

\begin{figure}
    \centering
    \hspace*{-1.3cm}
    \includegraphics[trim={0 4cm 0 0.0cm},scale=1.1]{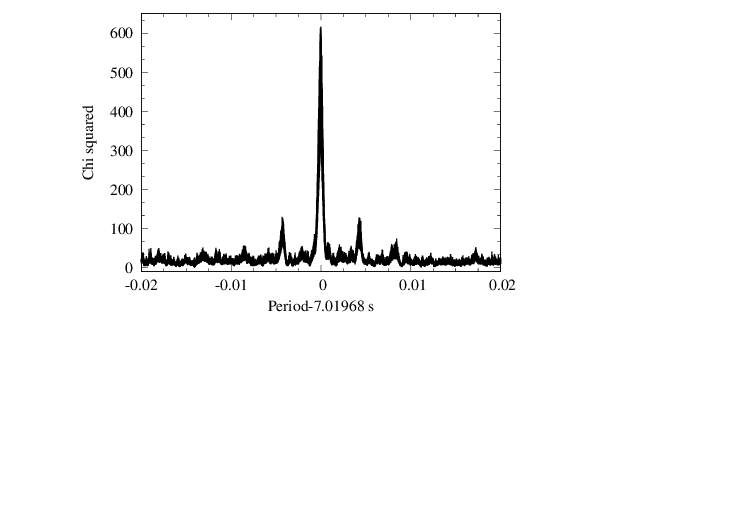}
    \caption{The Chi-square distribution plot obtained using the \textit{efsearch} method on {\it NuSTAR} light curve of  RX~J0032.9-7348, with the maximum Chi-square observed at a period of approximately 7.01968 s. 10$^{6}$ periods have been searched at a resolution of 10$^{-5}$ s.}
    \label{fig:7p02_efsearch}
\end{figure}

We searched for pulsations in the {\it NuSTAR} and {\it NICER} light curves using the \texttt{powspec} tool from the XRONOS package. We also used the power spectrum tool of \texttt{Stingray} package~\citep{2019ApJ...881...39H} and searched for pulsations over a broad frequency range using both methods. The resulting power density spectra (PDS) from the \texttt{Stingray} package are shown in Figure~\ref{fig:powspec}. For {\it NICER} observations, sharp peaks are observed only in observation ID~7205010102 (hereafter ID~102) and not in ID~7205010113 (hereafter ID~113). The peak at approximately 0.14 Hz ($\sim 7.02$ s) corresponds to the fundamental frequency, representing the rotational period of the pulsar, while the peak at approximately 0.28 Hz ($\sim 3.52$ s) is its first harmonic. We then applied the \texttt{efsearch} task from the FTOOL package to the light curves to determine the periodicity using the Chi-square maximization technique. Starting with an initial guess of 7.02 s, we searched for 10$^{6}$ periods at a resolution of 10$^{-5}$~s. The result obtained from the \texttt{efsearch} task for the {\it NuSTAR} observation is shown in Figure~\ref{fig:7p02_efsearch}. The pulse periods derived from the {\it NuSTAR} and {\it NICER} observations using \texttt{efsearch} are 7.0196 (1) s and 7.0243 (11) s, respectively, confirming the pulsations reported by \citet{2024ATel16901....1H}.  We estimated the period uncertainty through 1000 Monte Carlo simulations, where each light curve was generated by resampling the observed counts from a Poisson distribution and repeating the epoch-folding analysis, with the final error taken as the standard deviation of the resulting period distribution.

\begin{figure}
    \centering
    \hspace*{-0.7cm}
    \includegraphics[trim={0.0cm 2.0cm 0.0cm 0.0cm},scale=1.2]{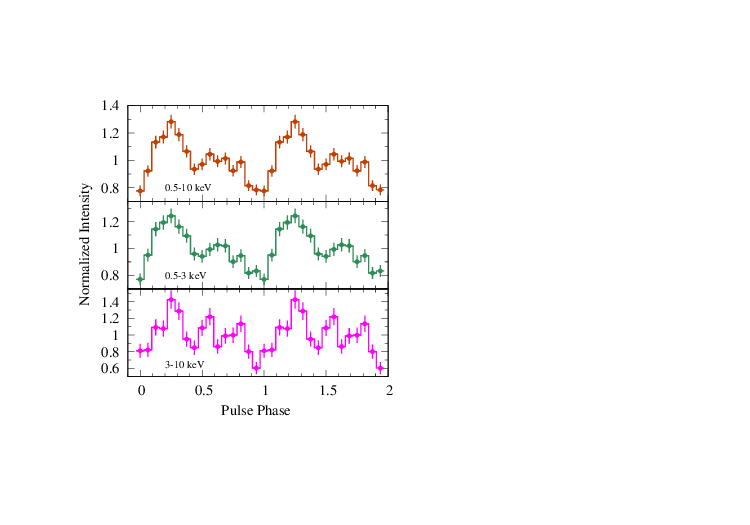}
    \caption{Energy-averaged and energy-resolved pulse profiles for {\it NICER} observation on MJD 60617 (ID~102) are presented where the pulse profiles are obtained in the energy bands of 0.5-10 keV, 0.5-3.0 keV, and 3-10 keV displayed from top to bottom, respectively.}
    \label{fig:nicer_pp}
\end{figure}

\begin{figure}
    \centering
    \hspace*{-0.7cm}
    \includegraphics[trim={0.0cm 1.5cm 0.0cm 0.0cm},scale=1.2]{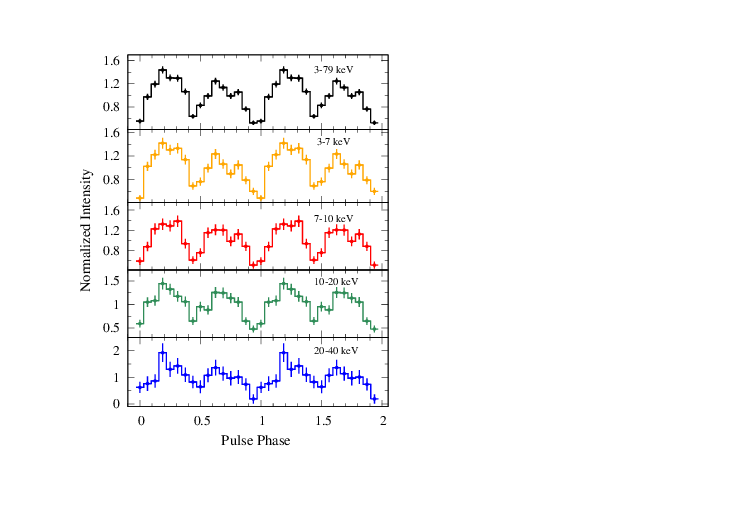}
    \caption{Energy-averaged and energy-resolved pulse profiles are generated from background-subtracted light curves for {\it NuSTAR} observation on MJD 60631 are presented where the pulse profiles are obtained in the energy bands of 3-79 keV, 3-7 keV, 7-10 keV,  10-20 keV, and 20-40 keV displayed from top to bottom, respectively.}
    \label{fig:nuEreolved_pp}
\end{figure}

Subsequently, we analyzed the energy-averaged and energy-resolved pulse profiles to probe the geometry of the emission region and its energy dependence. Each light curve is folded at the corresponding spin period of the pulsar utilizing the \texttt{efold} task within the FTOOLS package. To check the emission geometry in the soft X-ray regime, we initially analyzed the pulse profile of {\it NICER} observation (see Figure~\ref{fig:nicer_pp}). The pulse profile obtained in the $0.5-10$ keV energy range is double-peaked and asymmetric in nature. The pulse profile obtained in the 0.5-3 keV energy band is similar in nature. However, in the 3-10 keV pulse profile, an absorption dip is present at $\approx$0.7 phase, and a phase shift of $\approx$0.05 is also observed.  Then, we analyzed the pulse profile using the {\it NuSTAR} light curve to understand its behavior in the hard X-ray regime. The pulse profile in the 3–79 keV range is asymmetric and exhibits a double-peaked structure. Energy-resolved pulse profiles were generated for the 3–7, 7-10, 10–20, and 20–40 keV bands, as illustrated in Figure~\ref{fig:nuEreolved_pp}. While the double-peaked nature of the pulse profile persists across these energy bands, the relative intensities of the peaks vary with energy. As data in the 40-79 keV range are background-dominated, pulsations are not detected in this energy range.

Furthermore, we calculated the pulsed fraction (PF) of the pulse profiles to quantify the amount of pulsed emission from the source and its variation with energy using the root mean square (RMS) method given by: 
\begin{equation}
   PF = \frac{(\sum_{i=1}^{N} (r_{i}-\overline{r})^{2}/N)^{1/2}}{\overline{r}}
   \label{eq:1}
\end{equation}

\begin{figure}
    \hspace*{-0.0cm}
    \includegraphics[trim={0 2.2cm 0 0.7cm},scale=0.9]{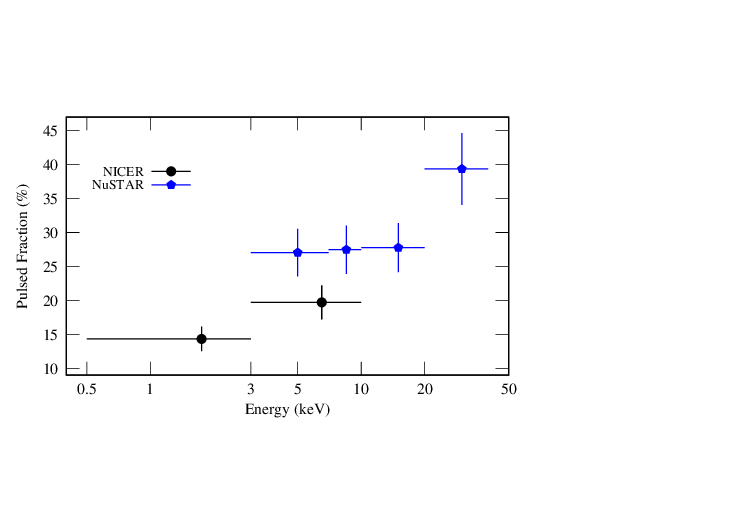}
    \caption{Energy-dependent evolution of the pulsed fraction (PF). The PF measurements from {\it NICER} are shown in black and those from {\it NuSTAR} in blue.}
    \label{fig:pf_energy}
\end{figure}

Where $r_{i}$ is the count rate in the $i$th phase bin of the pulse profile, $\overline{r}$ is the average count rate, and $N$ is the total number of phase bins. The variation of pulsed fraction (PF) with energy is depicted in Figure~\ref{fig:pf_energy}. For the {\it NICER} observation (ID~102), the PF in the 0.5–10 keV energy range is estimated as 14.6 ($\pm$1.8)\%. It increases from 14.3 ($\pm$1.8)\% in the 0.5–3 keV band to 19.7 ($\pm$2.5)\% in the 3–10 keV band. No strong pulsations were detected from NICER observation (ID~113; see middle panel of Figure~\ref{fig:powspec}). However, we folded the corresponding light curve, assuming a period within the range of 6.9 to 7.2 seconds. The PF varied between 9-21\%, suggesting an upper limit on the PF of $<$21\%, which may indicate a weak pulsation. In the {\it NuSTAR} data, the PF in the 3–79 keV range is 27.0 ($\pm$3.5)\%, remaining around 27\% in the 3–20 keV band and then increasing to 39 ($\pm$6)\% in the 20–40 keV band.

\section{Spectral Analysis and Results}
\label{sec:xray_spectral}

\subsection{Phase-averaged spectroscopy}
The X-ray spectral analysis is performed using XSPEC v-12.14.1 \citep{1996ASPC..101...17A} package. The spectral analysis for the {\it NICER} and {\it NuSTAR} observations are carried out in the $0.5-8$ keV and 3-50 keV energy ranges, respectively. The data beyond 8 keV and 50 keV are background dominated in the case of {\it NICER} and {\it NuSTAR} observations, respectively. Spectra are binned to ensure 20 counts per bin to facilitate the $\chi^{2}$ statistics in our analysis. For assessing the line-of-sight X-ray absorption, we utilize the \texttt{wilm} abundance table \citep{2000ApJ...542..914W}  alongside \texttt{Vern} photo-ionization cross section \citep{1996ApJ...465..487V}. The uncertainties on the parameters are calculated within the 90\% confidence range.

\begin{figure}
\centering
\hspace*{-1.5cm}
\includegraphics[trim={0 2.0cm 0 0.6cm},scale=1.2]{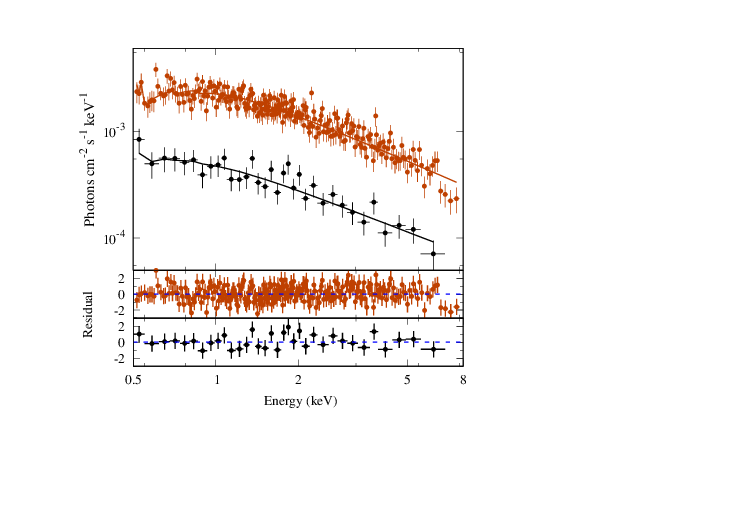}
\caption{The 0.5–8 keV {\it NICER} spectra and best-fit absorbed power-law model for observation IDs 102 (brown) and 113 (black) are shown in the top panel. The second and third panels display the residuals for IDs 102 and 113, respectively, after applying the best-fit model.}
\label{fig:rx_nicer_pl}
\end{figure}

\begin{figure}
\centering
\hspace*{-0.6cm}\includegraphics[trim={0 1cm 0 1.5cm},scale=0.38, angle=-90]{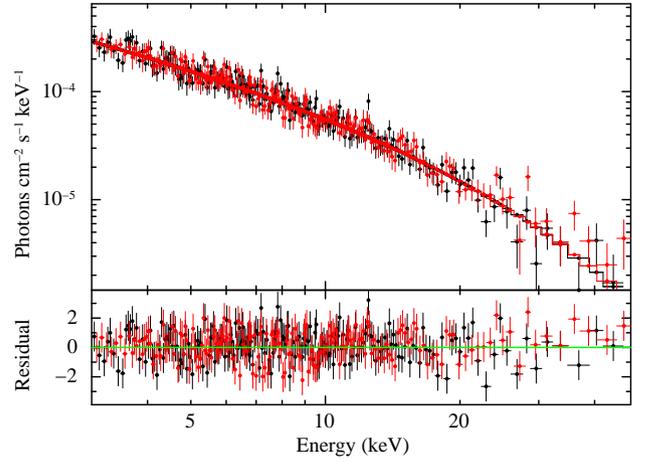}
\caption{The 3-50 keV {\it NuSTAR} source spectra and the best-fit absorbed cutoff power-law model spectra are shown in the top panel. The bottom panel presents the residuals after applying the best-fit model.}
\label{fig:rx_cutoff}
\end{figure}

The $0.5-8$ keV {\it NICER} spectra are best fitted with an absorbed power-law model: \texttt{Tbabs(powerlaw)}. The source spectra, the best-fitted model, and the corresponding residuals for observation ID~102 and ID~113 are shown in Figure~\ref{fig:rx_nicer_pl}. The values of the spectral parameters are shown in Table~\ref{tab:specfit_nicer}. Similarly, we performed spectral analysis on the {\it NuSTAR} observation in the $3-50$ keV energy range. At first, we modeled the spectra with the absorbed power-law model: \texttt{constant(Tbabs$\times$powerlaw)}. While fitting the {\it NuSTAR} spectra, we fixed the value of the line-of-sight equivalent hydrogen column density ($N_{\text{H}}$) to the Galactic value, as {\it NuSTAR} does not cover energies below 3 keV, where Galactic absorption is significant. The Galactic $N_{\text{H}}$ value for RX~J0032.9-7348 is 2.25$\times$10$^{21}$ cm$^{-2}$ \citep{2016A&A...594A.116H}. The absorbed power-law model did not adequately fit the high-energy part of the spectrum. We then used the absorbed power-law model with high energy exponential cutoff: \texttt{constant(Tbabs$\times$cutoffpl)}, which provided a good fit to the spectra, as shown in Figure~\ref{fig:rx_cutoff}. The best-fit parameters obtained from the spectral fitting are presented in Table~\ref{tab:specfit}.

\begin{table*}
\caption{Best-fit parameters obtained from the spectral analysis of {\it NICER} observations of RX~J0032.9-7348 in the $0.5-8$ keV energy range using an absorbed power-law model.}
 \label{tab:specfit_nicer}
\centering
\begin{tabular}{ llll } 
\hline
MJD               &                                            & 60617                      & 60633                       \\
\hline
Model             & Parameters                                 &        ID~102              & ID~113	       			\\  [4 pt]
\hline  
TBabs             & $\rm N_{H} (10^{22}~cm^{-2}$)              & $0.17_{-0.02}^{+0.02}$     & $0.09_{-0.06}^{+0.06}$      \\ [6 pt]
Powerlaw          & Photon Index ($\Gamma$)                    & $1.08_{-0.05}^{+0.05}$     & $0.97_{-0.19}^{+0.20}$      \\ [6 pt]
                  & Normalization (10$^{-3}$)                  & $3.02_{-0.02}^{+0.02}$     & $0.56_{-0.10}^{+0.13}$      \\ [6 pt]
Unabsorbed Flux ($0.5-50$ keV) & ($\rm 10^{-11}~erg~s^{-1}~cm^{-2}$) & $7.88_{-0.49}^{+0.50}$ & $1.78_{-0.40}^{+0.50}$      \\ [4 pt] 
Fit Statistics    & $\rm \chi_{red}^{2}/d.o.f.$                & 1.02/242                   & 0.77/31                    \\
\hline
\end{tabular}
\end{table*}

\begin{table}
\setlength{\tabcolsep}{5pt}
\caption{Best-fit parameters obtained from the spectral analysis of {\it NuSTAR} observation of RX~J0032.9-7348 in the 3-50 keV energy range using an absorbed cutoff power-law model.}
 \label{tab:specfit}
\centering
\begin{tabular}{l@{\hspace{0.5cm}}ll} 
\hline
Models                      &Parameters                           &Cutoffpl                         \\  [4pt]
\hline
TBabs                       &$\rm N_{H} (10^{22}~cm^{-2}$)        &0.22 (fixed)                    \\  [6 pt]
Cutoffpl                    &Photon Index ($\Gamma$)              &$1.1^{+0.1}_{-0.1}$            \\  [6 pt]
                            &$\rm E_{cut}$ (keV)                  &$17.1^{+3.7}_{-2.7}$            \\  [6 pt]
                            &Normalization (10$^{-3}$)            &$1.2^{+0.2}_{-0.1}$            \\  [10pt]
Unabsorbed Flux ($0.5-50$ keV) &($\rm 10^{-11}~erg~s^{-1}~cm^{-2}$)  &$2.32^{+0.08}_{-0.09}$        \\  [6 pt] 
Fit Statistics              &$\rm\chi_{red}^{2}/d.o.f.$           &1.002/386                     \\  [6 pt]
\hline \\
\end{tabular}\\

\end{table}

\begin{figure}
    \centering
    \hspace*{-0.5cm}
    \includegraphics[trim={0.0cm 0.8cm 0.0cm 0.0cm},scale=1.0]{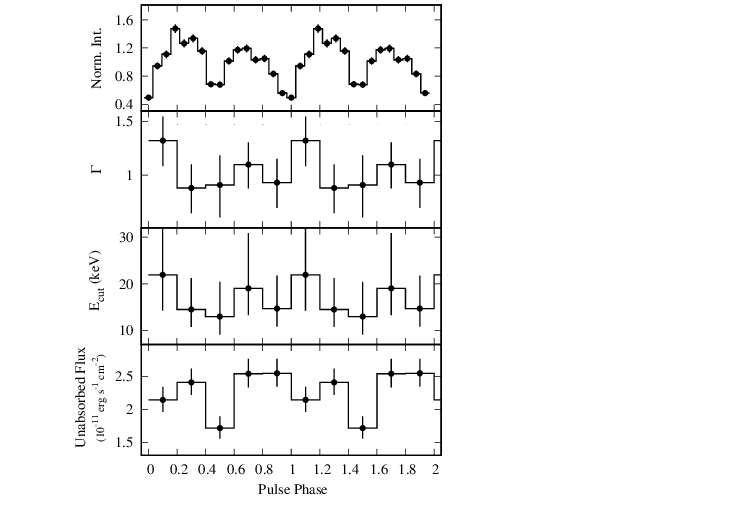}
    \caption{Spectral parameters obtained from the phase-resolved spectroscopy of RX~J0032.9-7348 using {\it NuSTAR} observation with absorbed cutoff power-law model. The top panel shows the pulse profile in the 3–79 keV energy range, while the 2nd, 3rd, and 4th panels display the phase variations of the photon index, cutoff energy, and unabsorbed flux in the 0.5–50 keV energy range, respectively. The uncertainties associated with the parameters are calculated within the 90\% confidence range.}
    \label{fig:pr_cutoffpl}
\end{figure}

\subsection{Phase-resolved spectroscopy}
We performed phase-resolved spectroscopy using {\it NuSTAR} observation to investigate phase-dependent variations in the spectral parameters derived from the absorbed cutoff power-law model. Phase-resolved spectra were extracted using the \texttt{XSELECT} package for phase intervals of 0.0–0.2, 0.2–0.4, 0.4–0.6, 0.6–0.8, and 0.8–1.0. As in the phase-averaged analysis, N$_{\text{H}}$ was fixed at 2.25$\times$10$^{21}$ cm$^{-2}$. Spectral fitting was carried out in the 3–50 keV energy range, and phase-dependent variations of the best-fit parameters are shown in Figure~\ref{fig:pr_cutoffpl}. The uncertainties associated with the parameters are calculated within the 90\% confidence range.  From Figure~\ref{fig:pr_cutoffpl}, the parameter variations appear within the uncertainty limits due to relatively large error bars. To evaluate the statistical significance of the variations in photon index ($\Gamma$) and cutoff energy (E$_{\text{cut}}$), we conducted a chi-square test. The test yielded p-values of 0.90 for $\Gamma$ and 0.86 for E$_{\text{cut}}$, suggesting that the observed variations are not statistically significant. Within the current statistical uncertainties, $\Gamma$ and  E$_{\text{cut}}$ show no significant variation across the different phase bins.

\section{Discussion}
\label{discussion}
We studied the timing and spectral properties of the unexplored X-ray transient source RX~J0032.9-7348 in the SMC. Using {\it NICER} and {\it NuSTAR} observations in November 2024, we present its broadband characteristics for the first time. Detection of strong X-ray pulsations, with a period of $\approx$7.02 s (see Figure~\ref{fig:7p02_efsearch}), confirms its nature as an X-ray pulsar. Our findings align with the results obtained from the XMM-Newton observations \citep{2024ATel16901....1H}, establishing RX~J0032.9-7348 (= SXP 7.02) as a newly identified X-ray pulsar in the SMC. Additionally, we analyzed the spin period evolution of RX~J0032.9-7348 using data from {\it NuSTAR} and {\it NICER}. Between MJD 60617 and 60631, the spin period decreased at a rate of $-$(3.3$\pm$0.8)$\times10^{-4}$ s day$^{-1}$, indicating an increase in spin frequency. This spin-up suggests that the pulsar gained angular momentum, likely due to accretion processes during the X-ray brightening phase. While the orbital period of the binary is unknown, the $\approx$7 s pulsation implies a 20–30 d orbital period based on the Corbet diagram of known SMC pulsars~\citep{2017AAS...22923305Y}.

Furthermore, we studied energy-averaged and energy-resolved pulse profiles in broadband energy regimes. Broadband study helps in understanding the geometry of the X-ray emitting region, its variation with energy, and the distribution of plasma in the magnetosphere. The shape of pulse profiles also reveals important information about the mass accretion rate as it varies with changes in the rate.  The {\it NICER} pulse profile obtained at a luminosity $\approx$3.6$\times$10$^{37}$ erg s$^{-1}$ are asymmetric and double-peaked and the pulse profile shape exhibit moderate variation with energy (see Figure~\ref{fig:nicer_pp}). A significant absorption dip is present at the 0.7 phase only in the 3-10 keV pulse profile. The 3-79 keV {\it NuSTAR} pulse profile, obtained at a luminosity of $\approx$1.1$\times$10$^{37}$ erg s$^{-1}$, is also double-peaked and asymmetric in nature (see Figure~\ref{fig:nuEreolved_pp}). A moderate energy-dependent variation in the pulse profile is observed. Unlike the pulse profiles from {\it NICER}, there is no significant absorption dip in the {\it NuSTAR} pulse profiles. These pulse profiles remain double-peaked across the 3-40 keV range. Notably, a phase shift is observed in the 20-40 keV range.

The absorption dip observed in the {\it NICER} pulse profiles at a luminosity level of $\approx$3.6$\times$10$^{37}$ erg s$^{-1}$, near the 0.7 phase, could be due to the absorption of low energy photons by accreting material locked in the magnetosphere asymmetrically \citep{1983ApJ...270..711W}. Dips in the pulse profile due to absorption of radiation from the pulsar by matter stream in a specific phase of the magnetosphere is also observed in various other Be/X-ray binary pulsars such as V0332+53, 1A~0535+262, EXO~2030+375, GX~304-1, and RX~J0209.6-7427 \citep{2006MNRAS.371...19T, 2008ApJ...672..516N,2013ApJ...764..158N, 2015RAA....15..537N, Epili2017MNRAS.472.3455E, 2016MNRAS.457.2749J, 2020MNRAS.494.5350V,2024ApJ...963..132C}.  Interestingly, we did not observe any absorption dips in the pulse profile at a lower luminosity level of $\approx$1.1$\times$10$^{37}$ erg s$^{-1}$.

We examined the pulsed fraction (PF) and its energy dependence using {\it NICER} and {\it NuSTAR} observations (see Figure~\ref{fig:pf_energy}). For {\it NICER} observations at a luminosity of $\approx$3.6$\times$10$^{37}$ erg s$^{-1}$, the PF varied from  14\% to 19\% across the 0.5–10 keV range, exhibiting a weak energy dependence. In contrast, {\it NuSTAR} observations at a lower luminosity of $\approx$1.1$\times$10$^{37}$ erg s$^{-1}$ showed a  higher PF of around 27\%, with no significant variation in 3-20 keV energy range and increases to a value of 39 ($\pm$6)\% in 20-40 keV energy range. The increase in PF with energy suggests that the hard X-ray photons originate from a more compact region~\citep{1997ApJS..113..367B,2018MNRAS.474.4432J,2021MNRAS.500..565B,2023MNRAS.521.3951J}.

We carried out a broadband spectral analysis to understand the physical mechanism of X-ray emission from RX~J0032.9-7348 using {\it NICER} and {\it NuSTAR} observations. The {\it NICER} spectra are described well with an absorbed power-law model. The best-fitted values of the spectral parameters are presented in Table~\ref{tab:specfit_nicer}. The {\it NICER} observations were taken on MJD 60617 and 60633 and during the span of 16 days, the luminosity of the source decreases from 3.6$\times$10$^{37}$ erg s$^{-1}$ to 8.2$\times$10$^{36}$ erg s$^{-1}$. The luminosity for both {\it NICER} and {\it NuSTAR} observations is computed within the $0.5-50$ keV energy range to maintain coherence and enable consistent comparison of source properties. The calculation assumes a source distance of 62 kpc, the average distance to the SMC~\citep{2014ApJ...780...59G}. The luminosity observed during the {\it NuSTAR} and {\it NICER} observations is an order of magnitude higher than that measured by ROSAT~\citep{1996A&A...312..919K}. The value of the photon index, though, did not vary much and remained around 1.0 during this time span. The value of the line-of-sight column density (N$_{\text{H}}$) obtained to be around 10$^{21}$ cm$^{-2}$. The {\it NuSTAR} spectral analysis is carried out in the 3-50 keV energy range are well described with an absorbed cutoff power-law model. The best-fitted values of the parameters are presented in Table~\ref{tab:specfit}. An absorbed cutoff power-law model is typically used to explain the broadband spectrum of the accretion-powered pulsar~\citep{2002ApJ...580..394C}. During the {\it NuSTAR} observation, the source was emitting at a luminosity level of 1.1$\times$10$^{37}$ erg s$^{-1}$. The value of the photon index ($\Gamma$) and the cutoff energy (E$_{\text{cut}}$) are obtained to be   $1.1_{-0.1}^{+0.1}$ and 17.1$_{-2.7}^{+3.7}$ keV, respectively. The $\Gamma$ and E$_{\text{cut}}$ obtained from {\it NICER} and {\it NuSTAR} spectral analysis are constrained between 0.5–1.5 keV and 10–30 keV, respectively, which is typical for accreting pulsars \citep{1983ApJ...270..711W}. We did not find the presence of an iron line in both {\it NICER} and {\it NuSTAR} spectra. No evidence of a cyclotron resonant scattering feature (CRSF) signature was found in {\it NuSTAR} spectra.

We did not detect any pulsations in the {\it NICER} light curve observed on MJD 60633 using both power density spectra (PDS) analysis and the epoch-folding methods. During this observation, the source was emitting at a luminosity level of 8.2$\times$10$^{36}$ erg s$^{-1}$, in the $0.5-50$ keV energy range. This suggests that the source may be close to the propeller regime ~\citep{1975A&A....39..185I}. However, caution is needed as the non-detection of pulsations during the second {\it NICER} observation may also result from the lower exposure time. Historically, the minimum intensity level at which the source was reported by ~\citet{1996A&A...312..919K} from the Rosat All-Sky Survey (RASS) in 1993 at a flux level of $\approx$2$\times$10$^{-12}$ erg cm$^{-2}$s$^{-1}$ in the 0.2–2 keV energy range, corresponding to a luminosity of 9.2$\times$10$^{35}$ erg s$^{-1}$. 
In the case of other SMC pulsars such as XTE~J0119-731 (SXP~2.13), SMC~X-2 (SXP~2.37), and SMC~X-3 (SXP~7.78), the propeller regime has been detected around a luminosity of 1.8$\times$10$^{36}$ erg s$^{-1}$~\citep{2017MNRAS.470.1971V}, 4$\times$10$^{36}$~erg~s$^{-1}$~\citep{2017ApJ...834..209L}, and (0.3--7)$\times$10$^{35}$~erg~s$^{-1}$ ~\citep{2017A&A...605A..39T}, respectively.  
If we assume that the luminosity at which RX~J0032.9-7348 (SXP~7.02) enters the propeller state somewhere between 10$^{34}$ and 5$\times$10$^{36}$ erg s$^{-1}$ then using the equation given in \citet{2002ApJ...580..389C}, the magnetic field of the neutron star can be calculated as follows:

\begin{equation}
    B_{12} \simeq ( L_{lim} \times 2.5 \times10^{-38} k^{-7/2} P^{7/3}  M_{1.4}^{2/3}  R_{6}^{-5} )^{1/2}~~Gauss,
    \label{eq:mag}
\end{equation}

where, $L_{lim}$ is the limiting luminosity for the onset of the propeller state, $R_6$ denotes the neutron star radius in units of $10^6$ cm, $M_{1.4}$ represents the neutron star mass in units of 1.4$M_\odot$, $P$ is the spin period (in seconds) of the pulsar, and $B_{12}$ refers to the surface magnetic field strength in units of 10$^{12}$~G, assuming a dipole magnetic field configuration. The parameter $k$ relates the magnetospheric radius in the case of disc accretion to the Alfvén radius for spherical accretion ($R_m = kR_A$) and is commonly taken as $k = 0.5$ \citep{1978ApJ...223L..83G}.

Using the values $L_{\text{lim}}$ = $10^{34}$--5$\times10^{36}$ erg s$^{-1}$, P = 7.02 s, neutron star mass = 1.4 $ M_\odot$, and neutron star radius = $10^6$ cm, the magnetic field is calculated in a range between 1.4$\times 10^{11}$ and 3.2$\times 10^{12}$~G. The most efficient way to constrain the magnetic field of accreting X-ray pulsars is through the detection of CRSF; however, we did not detect any signatures of the CRSF in the 3-50 keV {\it NuSTAR} spectra. The absence of the CRSF might be attributed to the low luminosity state of the source, and future bright state observations of the source will be crucial to constrain the magnetic field of the source. 

We conducted phase-resolved spectroscopy to investigate the rotational phase variation of emission mechanisms in RX~J0032.9-7348. Figure~\ref{fig:pr_cutoffpl}, shows variations in the photon index ($\Gamma$) and cutoff energy (E$_{\text{cut}}$) across different pulse phases. However, statistical analysis indicates no significant variation in $\Gamma$ and E${_\text{cut}}$ with phase. This uniformity implies that the physical conditions governing the X-ray emission, such as the geometry and properties of the emission region, do not change appreciably with phase, or that any phase-dependent variation is well within the current error bars.  

\section{Conclusion}
\label{conclusion}
We studied the newly discovered SMC X-ray pulsar RX~J0032.9-7348 in broadband X-ray energy range using {\it NuSTAR} and {\it NICER} during its X-ray brightening state in November 2024. Our timing analysis identified a pulsation period of approximately 7.02 s in the light curve. The pulse profile across a broad energy range was found to be double-peaked, with moderate variations observed as a function of energy. Spectral analysis indicated that the $0.5-8$ keV {\it NICER} spectrum is well described by an absorbed power-law model, while the 3-50 keV {\it NuSTAR} spectrum is best fitted with an absorbed power-law model modified by a high-energy cutoff. No evidence of iron emission line or cyclotron resonance scattering features was found in the energy spectrum. During our observation time span, the luminosity of the source varied between 8.2$\times$10$^{36}$ and 3.7$\times$10$^{37}$~erg s$^{-1}$. Considering the transition of the source to the propeller regime, we suggest the magnetic field of the source to be between 1.4$\times$10$^{11}$--3.2$\times$10$^{12}$~G, assuming a limiting luminosity of $10^{34}$--5$\times10^{36}$ erg s$^{-1}$.

\section*{Acknowledgements}
We thank the anonymous reviewer for the suggestions, which helped us to improve the manuscript. The research work at the Physical Research Laboratory is funded by the Department of Space, Government of India. AJ acknowledges support from FONDECYT Postodoctoral fellowship (3230303). This research has made use of {\it NuSTAR}, {\it NICER} mission data, and X-ray data analysis software provided by the High Energy Astrophysics Science Archive Research Center (HEASARC), which is a service of the Astrophysics Science Division at NASA/GSFC. We thank the {\it NuSTAR} SOC Team for making this ToO observation possible.

\section*{Data Availability}
The X-ray data from {\it NuSTAR} and {\it NICER} are publicly available and can be accessed from HEASARC Archive (\texttt{heasarc.gsfc.nasa.gov/db-perl/W3Browse/w3browse.pl}).






\bibliographystyle{mnras}
\bibliography{rxj0032} 



\bsp	
\label{lastpage}
\end{document}